\documentclass[twocolumn,english,aps,prb,groupedaddress,showpacs,superscriptaddress,nobibnotes]{revtex4-1}
\usepackage{amsmath,amssymb,amsthm}
\usepackage[T1]{fontenc}

\usepackage{natbib}
\usepackage{mathbbol,amsmath,amsfonts,amssymb,bm,color,dsfont,bbold}
\usepackage{graphicx,psfrag,array,braket}
\usepackage{float}
\usepackage{soul}
\allowdisplaybreaks

\usepackage{color}
\usepackage{verbatim}
\usepackage{esint}
\usepackage{soul}
\usepackage[normalem]{ulem}

\makeatletter

\usepackage{caption}
\usepackage{subcaption}
\usepackage[dvipsnames,svgnames,x11names]{xcolor} 
\usepackage[pdftex]{hyperref}
\hypersetup{
    breaklinks=true,
    colorlinks=true,
    linkcolor=MidnightBlue,
    urlcolor=NavyBlue,
    citecolor=Blue,
    pdftitle={ },
    pdfauthor={ },
}

\usepackage{wrapfig}
\usepackage{braket}
\usepackage{array}
\usepackage{relsize}
\usepackage{indentfirst}
\usepackage{simplewick}
\usepackage{mathrsfs}

\usepackage{orcidlink}

\makeatother

\usepackage{lipsum}

\providecommand{\ignore}[1]{}
\providecommand{\aucmnt}[1]{#1}
\def\be{\begin{equation}}
\def\ee{\end{equation}}

\renewcommand{\aucmnt}[1]{}
\newcommand{\Comment}[1]{}
\newcommand{\Eq}[1]{Eq.~(\ref{#1})}

\begin{document}
\title{Second-quantized approach to the study of Halperin state in fractional quantum Hall effect}
\author{Li Chen\,\orcidlink{0000-0002-4870-9065}}
\affiliation{College of Physics and Electronic Science, Hubei Normal University, Huangshi 435002, China}\affiliation{Hubei Key Laboratory of Photoelectric Materials and Devices, Hubei Normal University, Huangshi 435002, China}
\author{Zhiping Yao}
\affiliation{College of Physics and Electronic Science, Hubei Normal University, Huangshi 435002, China}

\date{\today}
\begin{abstract}
We give a recursion relation for the second-quantized fermionic (bosonic) Halperin state, which avoids exact diagonalization of its two-component first-quantized parent Hamiltonian. We validate this formula by proving that the \textit{second-quantized} Halperin state, as recursively defined in this formula, is indeed a zero mode of the corresponding second-quantized parent Hamiltonian and that it has the correct filling factor.
\end{abstract}
\maketitle

\section{Introduction}
The study of strongly correlated systems has been a focus of condensed  matter physics for decades,  of which fractional quantum Hall (FQH) system\cite{RevModPhys.71.S298,RevModPhys.75.1101,compositeReview,QHCFTRevMod} has been a prototype. FQH system possess exotic properties such as fractional charge and  (non)Abelian exchange statistics for its quasiparticle/quasihole excitation\cite{jainbook,RevModPhys.80.1083}, or even the graviton-like excitation\cite{PhysRevLett.123.146801,PhysRevResearch.3.023040,PhysRevB.104.L121106,graviton2024,YANGInnovation}. 

Most of many-body electronic states studied in fractional quantum Hall effect are single-component, if the Landau level mixing can be neglected and the Coulomb repulsion dominates the Zeeman splitting. 
Prominent examples are Laughlin state\cite{Laughlin}, Jain composite fermion state\cite{jainbook,compositeReview}, Pfaffian state\cite{MR}, Read-Rezayi state,\cite{RR} etc.  However, multi-component FQH states may also arise in systems involving extra degrees of the freedom, e.g. spin, valley, layer, etc.  Prominent examples include extensively studied Halperin state\cite{Halperin}, Haldane-Rezayi state\cite{HaldaneRezayi}, among others\cite{FQHPlasmaSU,SY1,SpinQHsqueez,PhysRevB.90.165102,HalperinMPS,PhysRevB.100.245122,Zeng,HalperinChernband}. 

The above FQH states are mainly studied with their wave functions\cite{wf}satisfying certain clustering property when two or multiple particles coalesce. This clustering property in turn lead to the construction of their model Hamiltonians. On the other hand, the second-quantized approach has also been emerging for FQH states, which has been successful in constructing parent Hamiltonians for FQH states involving multiple Landau levels. In that case, the wave functions for FQH states have  clustering property too complicated to guarantee a successful searching of the  parent Hamiltonian in the first-quantized approach. Indeed, the second-quantized parent Hamiltonians have been constructed for unprojected Jain composite fermion states\cite{PhysRevLett.124.196803} and parton states\cite{PartonFQH}, which reveal the emergent SU$(n)$ symmetry of their zero mode.

In addition, the second-quantized approach has also been used to construct the recursion relations and nonlocal order parameters for Laughlin state\cite{Chen14}, unprojected Jain composite fermion state\cite{Chen19}, and Pfaffian state\cite{Pfaffian2023} as well its particle-hole conjugate, the anti-Pfaffian state. 

In this work, we generalize this second-quantized approach to the case of two-component Halperin state. The paper is organized as follows.
In Sec. \ref{Hal} we introduce the Halperin state in the context of the first quantization.
In Sec. \ref{Halre} we give a  second-quantized recursion relation \eqref{rere} for the Halperin state, and then in Sec. \ref{zm123} prove the second-quantized Halperin state, as recursively defined in \Eq{rere}, satisfies all  zero mode conditions.
In Sec. \ref{RS} we prove that this state, as defined in the recursion relation, indeed has the correct  filling factor.
We present conclusion and discussion in Sec. \ref{condisc}.

\section{Second-quantized fermionic (bosonic) Halperin state}
\subsection{Halperin state and its parent Hamiltonian}\label{Hal}
The $2N$-particle Halperin $(m,m',n)$ state's first-quantized wave function\cite{Halperin} is  \begin{align}\label{1stHal}&\prod\limits_{1\leqslant i<j\leqslant N}\left(z_i-z_j\right)^m\prod\limits_{1\leqslant i<j\leqslant N}\left(w_i-w_j\right)^{m'}\notag\\&\times\prod\limits_{1\leqslant i,j \leqslant N}\left(z_i-w_j\right)^n.\end{align}
In the Halperin state, particles have internal degree of freedom, e.g. spin or valley, which can be symbolized as a layer index. The complex coordinate of the $i$-th particle in the first layer and that of the $j$-th particle in the second layer is thus labeled as $z_i$ and $w_j$, respectively. If particles in the first and second layer are fermions (particles in the first layer is still distinguishable from those in the second one), $m$ and $m'$ are odd; if particles in the first and second layer are bosons (particles in the first layer is still distinguishable from those in the second one), $m$ and $m'$ are even. We also employ stability criteria $m \geqslant n$ and $m' \geqslant n$ so that  $mm' \geqslant n^2$\cite{FQHPlasmaSU}. The case $m=m'=n$ corresponds to the Laughlin state with the extra degree of freedom. Hence we will focus on the case where $m > n$ and $m'>n$.

The two-body parent Hamiltonian for fermionic (bosonic) Halperin state is
\begin{align}\label{totalH} H&=\sum_{k<m}\partial_{z_i}^{k}\partial_{z^*_i}^{k}\delta(z_i-z_j)\delta(z^*_i-z^*_j)\notag\\&+\sum_{k<{m'}}\partial_{w_i}^{k}\partial_{w^*_i}^{k}\delta(w_i-w_j)\delta(w^*_i-w^*_j)\notag\\&+\sum_{k<n}\partial_{z_i}^{k}\partial_{z^*_i}^{k}\delta(z_i-w_j)\delta(z^*_i-w^*_j),\end{align} where  $\delta$ is the Dirac delta function, $z^*$ and $w^*$ are complex conjugate of respective coordinates.

In the approach of second-quantization, $ c_{i} $ is a fermionic (bosonic) operator which annihilates a fermion (boson)  of orbital  $\phi_r(z) =z^r/N_r$  in the lowest Landau level  (the single-particle wave function in genus-0 geometries such as disk, sphere and cylinder can be brought into this form).
We can introduce geometry-independent annihilation and creation operators $\bar{c}_r=c_r/N_r$ and $\bar{c}_r^\dagger=N_r c_r^\dagger$\cite{Chen19,2021exact} to treat disk, sphere and cylinder in the same framework\cite{ortiz}. For the sake of simplicity, we will not explicitly write down the bar for the annihilation and creation operators, although the reader should bear in mind that we are always referring to the geometry-independent operators, unless stated otherwise.

To distinguish between the two layers, from now on we will use the notations $a_r$ and $a_r^\dagger$  for fermionic (bosonic) geometry-independent annihilation and creation operators in the first layer,  $b_r$ and $b_r^\dagger$ for the second layer. The operators in the same layer obey the (anti) commutation relations for (fermions) bosons while two operators are always commutative if they belong to different layers. This will be used in the derivation of Eqs. \ref{Tpsi} and \ref{Vpsi}.

The above Hamiltonian in second quantization is of the following form,

\begin{align}\label{Ham} H=&\sum_{0  \leqslant k\leqslant m-1}\sum\limits_{R} T^{k\dag}_R  T^{k}_R+\sum_{0  \leqslant k\leqslant m'-1}\sum\limits_{R} U^{k\dag}_R  U^{k}_R\notag\\&+\sum_{\substack{0  \leqslant k\leqslant n-1}}\sum\limits_{R} V^{k\dag}_R  V^{k}_R,\end{align}
where the positive-semidefinite two-body operator $T^{k\dag}_R  T^{k}_R$,  $U^{k\dag}_R  U^{k}_R$ and  $V^{k\dag}_R  V^{k}_R$ projects onto a two-body state of relative angular momentum $k\hbar$ and total angular momentum $R\hbar$ in the lowest Landau level. The difference among them is that two-body state resides in the first and second layer in $T^{k\dag}_R  T^{k}_R$ and  $U^{k\dag}_R  U^{k}_R$, respectively.  By contrast, in the case of  $V^{k\dag}_R  V^{k}_R$, one particle is in the first layer and the other in the second one.

These operators have the form
\be\begin{split}
T^{k}_R=\sum_{\substack{0  \leqslant  i_1,i_2 \leqslant  R\\ i_1+i_2=R}}t_k(i_1,i_2) a_{i_2}a_{i_1},
\end{split} \ee
\be\begin{split}
U^{k}_R=\sum_{\substack{0  \leqslant  i_1,i_2 \leqslant  R\\ i_1+i_2=R}}u_k(i_1,i_2) b_{i_2}b_{i_1},
\end{split} \ee
\be\begin{split}
V^{k}_R=\sum_{\substack{0  \leqslant  i_1,i_2 \leqslant  R\\ i_1+i_2=R}}v_k(i_1,i_2) a_{i_2}b_{i_1},
\end{split} \ee
The form of $t_k$, $u_k$  and $v_k$ can be derived explicitly\cite{Simonetal2}. However,  in this paper it suffices to use the fact that all of them are polynomials of degree $k$ and that $t_k$ and  $u_k$ are (anti) symmetric polynomial in their variables for (fermion) boson.

Since the two-body operators in   the total Hamiltonian \eqref{Ham} are all positive-semidefinite, its zero mode $\ket{\psi_{\text{zero}}}$  must satisfy zero mode conditions \be\label{zo1} T^{k}_R  \ket{\psi_{\text{zero}}}=0,\ee \be\label{zo2} U^{k}_R  \ket{\psi_{\text{zero}}}=0,\ee and \be\label{zo3} V^{k}_R  \ket{\psi_{\text{zero}}}=0\ee for all $R$ and $k$ in \Eq{Ham}.

\subsection{Recursive formula for fermionic (bosonic) Halperin state}\label{Halre}
Now we write down the second-quantized recursion relation for Halperin $(m,m',n)$ state, which relates fermionic (bosonic) ($2N+2$)-particle Halperin state $\ket{\Psi_{2N+2}}$ to $2N$-particle Halperin state $\ket{\Psi_{2N}}$:
 \begin{align}\label{rere}
\ket{\Psi_{2N+2}}&= \frac{1}{N+1}\sum\limits_{l=0}^{n}(-1)^l {n \choose l} \sum_{k_1,k_2,k_3,k_4\geqslant 0}\, a_{n-l+k_1+k_2}^\dagger \notag\\&\times b_{l+k_3+k_4}^\dagger  S^{(m)}_{mN-k_1} S'^{(n)}_{nN-k_2}  S^{(n)}_{nN-k_3}S'^{(m')}_{m'N-k_4}  \ket{\Psi_{2N}},\end{align}
with the beginning of recursion $\ket{\Psi_{0}}=\ket{0}$, the vacuum.
Note that the above recursion relation can be derived from the first-quantized wave function, as proved in Appendix \ref{1stTo2nd}.
However, in a self-contained manner, we can also have this  recursion relation as our starting point, then (i) prove it satisfies all zero mode conditions Eqs.~\ref{zo1}, \ref{zo2}, and \ref{zo3};  (ii) prove it has the correct filling factor. These proofs will be the theme of the following sections.

The second-quantized flux-attachment operator $S^{(m)}_\ell $\cite{Chen14,Chen19}, which is closely related to $\prod_{i<j}\left(z_i-z_j\right)^m$ in the first quantization (see Appendix \ref{1stTo2nd}),  is defined as 
\be\label{Sm} \begin{split}
& {   S^{(m)}_\ell } = {( - 1)^\ell }\sum\limits_{i_1 + i_2+\cdots+i_m= \ell }  e_{i_1} e_{i_2}\cdots e_{i_m}\quad \text{for} \quad \ell\geqslant 0,\\&
   S^{(m)}_\ell=0\quad \text{for} \quad \ell<0.\end{split}\ee

The operator $e_i$ in the above definition is the second-quantized elementary-symmetric-polynomial operator which increases the angular momentum of each of $i$ fermions/bosons in the first layer by 1, 
\be\label{e}\begin{split}
  e_i =& \frac{1}{i!} \sum_{l_1,\dots ,l_i =  0}^{ + \infty } a_{l_1+1}^\dag a_{l_2+ 1}^\dag \cdots a_{l_i + 1}^\dag   a_{ l_i} \cdots a_{ l_2}a_{ l_1}  \\ &\text{for}\quad i>0,\\ e_0=&\mathbb{1},
\\ e_i=&0 \quad \text{for}\quad i<0.\end{split} \ee
Similarly, for the second layer we have
\be\label{Sprime} \begin{split}
& {   S'^{(m)}_\ell } = {( - 1)^\ell }\sum\limits_{i_1 + i_2+\cdots+i_m= \ell }  f_{i_1} f_{i_2}\cdots f_{i_m}\quad \text{for} \quad \ell\geqslant 0,\\&
   S'^{(m)}_\ell=0\quad \text{for} \quad \ell<0,\end{split}\ee
and \be\label{f}\begin{split}
  f_i =& \frac{1}{i!} \sum_{l_1,\dots ,l_i =  0}^{ + \infty } b_{l_1+1}^\dag b_{l_2+ 1}^\dag \cdots b_{l_i + 1}^\dag   b_{ l_i} \cdots b_{ l_2}b_{ l_1}  \\ &\text{for}\quad i>0,\\ f_0=&\mathbb{1},
\\ f_i=&0 \quad \text{for}\quad i<0.\end{split} \ee

These operators $S, S', e$ and $f$ have two important properties: (i) they give new zero mode when acting on an existing zero mode; (ii) they commute among themselves, independent of the nature of the particles involved.  These two properties derive from the fact that  elementary-symmetric-polynomial operator  can be expressed in terms of power-sum symmetric-polynomial operator by Newton-Girard relation\cite{Mazaheri14,Chen19}, and the fact that  the latter possesses these two properties\cite{Pfaffian2023}.  These two properties are necessary for the proofs in the following sections.

\subsection{Proof that recursively defined Halperin state satisfies zero mode conditions \eqref{zo1}, \eqref{zo2} and \eqref{zo3}}\label{zm123}

We now prove by the method of mathematical  induction that fermionic (bosonic) Halperin state, as recursively defined in \Eq{rere},
 is a zero mode of all $T^k_R$, $U^k_R$ and $V^k_R$.

{\textit{Proof.}} The beginning of mathematical induction consists of  $\ket{\Psi_{0}}=\ket{0}$,  $\ket{\Psi_{2}}$ and $\ket{\Psi_{4}}$. By using the recursion relation \Eq{rere}, we obtain
\begin{align}
\ket{\Psi_{2}}&= \sum\limits_{l=0}^{n}(-1)^l {n \choose l} \sum_{k_1,k_2,k_3,k_4}\, a_{n-l+k_1+k_2}^\dagger \notag\\&\times b_{l+k_3+k_4}^\dagger  S^{(m)}_{-k_1} S'^{(n)}_{-k_2}  S^{(n)}_{-k_3}S'^{(m')}_{-k_4}  \ket{0}.\end{align}
Note that $S'$ operator is the sum of products of $f$ operators, which have annihilation operators $b$ on the right, thus $S'^{(m')}_{-k_4}\ket{0}$ vanishes unless $-k_4=0$. The same applies to $S$.
We thus obtain 

\be\label{psi2}
\ket{\Psi_{2}}= \sum\limits_{l=0}^{n}(-1)^l {n \choose l} \, a_{n-l}^\dagger  b_{l}^\dagger   \ket{0}.\ee


By using the recursion relation \Eq{rere}, the second-quantized form of $\ket{\Psi_{4}}$ is 
\begin{widetext}
\begin{align}\label{psi_4}\ket{\Psi_{4}}=&\frac{(-1)^{m+m'}}{2}\sum\limits_{l_1,l_2=0}^n (-1)^{l_1+l_2} {n \choose l_1}{n \choose l_2} \sum\limits_{k_1=0}^m \sum\limits_{k_2=0}^n \sum\limits_{k_3=0}^n \sum\limits_{k_4=0}^{m'} (-1)^{k_1+k_2+k_3+k_4}{m \choose k_1}{n \choose k_2}{n \choose k_3}{m' \choose k_4} \notag\\&\times a_{n-l_1+k_1+k_2}^\dagger a_{2n+m-l_2-k_1-k_3}^\dagger   b_{l_1+k_3+k_4}^\dagger b_{m'+n+l_2-k_2-k_4}^\dagger  \ket{0},\end{align}\end{widetext}
where we have used the commutation relations\cite{Chen14,Chen19,Pfaffian2023}
\be\label{Sad}
[S^{(m)}_l,a_r^\dagger]=\sum_{p=1}^{m} (-1)^p {m \choose p} a_{r+p}^\dagger  S^{(m)}_{l-p} \ee and 
\be\label{Sbd}
[S'^{(m)}_l,b_r^\dagger]=\sum_{p=1}^{m} (-1)^p {m \choose p} b_{r+p}^\dagger  S'^{(m)}_{l-p} \ee 
to move $S$ and $S'$  to the right of $a^\dagger$ and $b^\dagger$, and have used the fact that $S^{(m)}_{k}\ket{0}$ and $S'^{(m)}_{k}\ket{0}$ vanish unless $k=0$. 

$\ket{\Psi_{0}}$,  $\ket{\Psi_{2}}$ and $\ket{\Psi_{4}}$ are annihilated by all  $T^k_R$, $U^k_R$ and $V^k_R$. This is obvious for $\ket{\Psi_{0}}$ since $\ket{\Psi_{0}}$ is vacuum. It is also obvious that  $\ket{\Psi_{2}}$ has one particle in each layer, hence is automatically annihilated by all  $T^k_R$ and  $U^k_R$, which are  two-body annihilation operators in each layer. We act $V^k_R$ on $\ket{\Psi_{2}}$ to obtain
\be
V^k_R\ket{\Psi_{2}}= \delta_{n,R}\sum\limits_{l=0}^{n}(-1)^l {n \choose l} v_k(l,n-l)=0\ee  by using a combinatorial identity\cite{Ruiz96} \be\label{cb} \sum_{l=0}^{n} (-1)^{l} {n \choose l}l^p=0 \quad \text{for } 0\leqslant p <n\, , \ee  since $ v_k$ in $V^k_R$  is a polynomial of degree $k$ with $k\leqslant n-1$.

Now we act $T^k_R$ on $\ket{\Psi_{4}}$ to obtain
\begin{widetext}\begin{align}&T^k_R\ket{\Psi_{4}}
\notag\\=&\sum_{\substack{0  \leqslant  i_1,i_2 \leqslant  R\\ i_1+i_2=R}}t_k(i_1,i_2) a_{i_2}a_{i_1}\frac{(-1)^{m+m'}}{2}\sum\limits_{l_1,l_2,k_1,k_2,k_3,k_4} (-1)^{l_1+l_2+k_1+k_2+k_3+k_4} {n \choose l_1}{n \choose l_2} {m \choose k_1}{n \choose k_2}{n \choose k_3}{m' \choose k_4} \notag\\&\times a_{n-l_1+k_1+k_2}^\dagger a_{2n+m-l_2-k_1-k_3}^\dagger   b_{l_1+k_3+k_4}^\dagger b_{m'+n+l_2-k_2-k_4}^\dagger  \ket{0}\notag
\\=&(-1)^{m+m'}\sum\limits_{l_1,l_2,k_2,k_3,k_4}\delta_{R,3n+m-l_1-l_2+k_2-k_3} (-1)^{l_1+l_2+k_2+k_3+k_4} {n \choose l_1}{n \choose l_2} {n \choose k_2}{n \choose k_3}{m' \choose k_4} \notag\\&\times  b_{l_1+k_3+k_4}^\dagger b_{m'+n+l_2-k_2-k_4}^\dagger  \Big[\sum\limits_{k_1=0}^m (-1)^{k_1}{m \choose k_1}t_k(n-l_1+k_1+k_2,2n+m-l_2-k_1-k_3)\Big] \ket{0},\end{align}\end{widetext}
 where the summation in the square  bracket  is zero by using \Eq{cb},  on account of  the fact that  $t_k$  in $T^k_R$ is a polynomial of degree $k$ with $k\leqslant m-1$.
Similarly, it is also easy to prove that  $U^k_R\ket{\Psi_{4}}=V^k_R\ket{\Psi_{4}}=0$.

Now we start mathematical induction and assume
\be\label{IndBegin}  T^k_R\ket{\Psi_{2N}}=U^k_R\ket{\Psi_{2N}}=V^k_R\ket{\Psi_{2N}}=0\ee
for all $R$ and corresponding $k$ (see \Eq{Ham}), with $N \geqslant  4$. Then we have
\begin{widetext}\begin{align}\label{Tpsi}
& (N+1)T^k_R\ket{\Psi_{2N+2}} \notag\\ =&  \sum_{\substack{0  \leqslant  i_1,i_2 \leqslant  R\\ i_1+i_2=R}}t_k(i_1,i_2) a_{i_2}a_{i_1} \sum\limits_{l=0}^{n}(-1)^l {n \choose l} \sum_{k_1,k_2,k_3,k_4}\, a_{n-l+k_1+k_2}^\dagger   b_{l+k_3+k_4}^\dagger  S^{(m)}_{mN-k_1} S'^{(n)}_{nN-k_2}  S^{(n)}_{nN-k_3}S'^{(m')}_{m'N-k_4}  \ket{\Psi_{2N}} \notag\\
=& 2 \sum_{\substack{0  \leqslant  i_1,i_2 \leqslant  R\\ i_1+i_2=R}}t_k(i_1,i_2) a_{i_2}  \sum\limits_{k_1,k_2,k_3,k_4}(-1)^{n-i_1+k_1+k_2} {n \choose i_1-k_1-k_2}\, b_{n-i_1+k_1+k_2+k_3+k_4}^\dagger  \notag
\\&\times S^{(m)}_{mN-k_1} S'^{(n)}_{nN-k_2}  S^{(n)}_{nN-k_3}S'^{(m')}_{m'N-k_4}  \ket{\Psi_{2N}}\notag\\
&+ \sum\limits_{l=0}^{n}(-1)^l {n \choose l} \sum_{k_1,k_2,k_3,k_4}\, a_{n-l+k_1+k_2}^\dagger   b_{l+k_3+k_4}^\dagger  T^k_R S^{(m)}_{mN-k_1} S'^{(n)}_{nN-k_2}  S^{(n)}_{nN-k_3}S'^{(m')}_{m'N-k_4}  \ket{\Psi_{2N}}\notag\\
=& 2 T^k_R\ket{\Psi_{2N+2}} ,
\end{align}\end{widetext}
where we have used \Eq{rere} in the first step, $a_{i_2}a_{i_1}a_{r}^\dagger =\delta_{r,i_1}a_{i_2}+(-1)^{m}\delta_{r,i_2}a_{i_1}+a_{r}^\dagger  a_{i_2}a_{i_1}$  and the (anti) symmetry of $t_k(i_1,i_2)$ in the second step, and \Eq{arpsi} in the third step. The (anti) commutativity for (fermions) bosons has been reflected in $(-1)^{m} $.  We have also used the identity $T^k_R S^{(m)}_{mN-k_1} S'^{(n)}_{nN-i_1+n-l+k_1}  S^{(n)}_{nN-k_3}S'^{(m')}_{m'N-k_4}  \ket{\Psi_{2N}}=0$, on account of the induction beginning \eqref{IndBegin} and the fact that $S$ and $S'$ are zero mode generators, as shown in the previous section.

We obtain $(N+1)T^k_R\ket{\Psi_{2N+2}}=2T^k_R\ket{\Psi_{2N+2}}$.
Therefore, if $\ket{\Psi_{2N}}$ is a zero mode of all  $T^k_R$ with  $2N  \geqslant  4 $, so will be $\ket{\Psi_{2N+2}}$. By mathematical induction, the second-quantized fermionic (bosonic) Halperin state, as recursively defined in \Eq{rere}, satisfies the first zero mode condition \Eq{zo1}.\hfill $\blacksquare$

Note that  $U^k_R$ consists only of annihilation operators in the second layer, just like $T^k_R$ consists only of annihilation operators in the first layer, so the above routine can be carried over to prove that the second-quantized Halperin state satisfies the second zero mode condition \Eq{zo2} as well.

We finally act $V^k_R$ on $\ket{\Psi_{2N+2}}$:
\begin{widetext}\begin{align}\label{Vpsi}
& (N+1)V^k_R\ket{\Psi_{2N+2}} \notag\\ =&  \sum_{\substack{0  \leqslant  i_1,i_2 \leqslant  R\\ i_1+i_2=R}}v_k(i_1,i_2) a_{i_2}b_{i_1} \sum\limits_{l=0}^{n}(-1)^l {n \choose l} \sum_{k_1,k_2,k_3,k_4}\, a_{n-l+k_1+k_2}^\dagger   b_{l+k_3+k_4}^\dagger  S^{(m)}_{mN-k_1} S'^{(n)}_{nN-k_2}  S^{(n)}_{nN-k_3}S'^{(m')}_{m'N-k_4}  \ket{\Psi_{2N}} \notag\\
=&-  \sum\limits_{k_1,k_2,k_3,k_4}\Big[\sum\limits_{l=0}^{n}(-1)^l {n \choose l}v_k(n-l+k_1+k_2,l+k_3+k_4)\Big] \delta_{R,n+k_1+k_2+k_3+k_4}  S^{(m)}_{mN-k_1} S'^{(n)}_{nN-k_2}  S^{(n)}_{nN-k_3}S'^{(m')}_{m'N-k_4}  \ket{\Psi_{2N}}\notag
\\& + \sum_{\substack{0  \leqslant  i_1,i_2 \leqslant  R\\ i_1+i_2=R}}v_k(i_1,i_2) a_{i_2} \sum\limits_{k_1,k_2,k_3,k_4}(-1)^{i_1-k_3-k_4} {n \choose i_1-k_3-k_4}\, a_{n-i_1+k_1+k_2+k_3+k_4}^\dagger  S^{(m)}_{mN-k_1} S'^{(n)}_{nN-k_2}  S^{(n)}_{nN-k_3}S'^{(m')}_{m'N-k_4}  \ket{\Psi_{2N}}\notag
\\& + \sum_{\substack{0  \leqslant  i_1,i_2 \leqslant  R\\ i_1+i_2=R}}v_k(i_1,i_2)b_{i_1}\sum\limits_{k_1,k_2,k_3,k_4}(-1)^{n-i_2+k_1+k_2} {n \choose i_2-k_1-k_2} \, b_{n-i_2+k_1+k_2+k_3+k_4}^\dagger  S^{(m)}_{mN-k_1} S'^{(n)}_{nN-k_2}  S^{(n)}_{nN-k_3}S'^{(m')}_{m'N-k_4}  \ket{\Psi_{2N}}\notag\\
&+ \sum\limits_{l=0}^{n}(-1)^l {n \choose l} \sum_{k_1,k_2,k_3,k_4}\, a_{n-l+k_1+k_2}^\dagger   b_{l+k_3+k_4}^\dagger  V^k_R S^{(m)}_{mN-k_1} S'^{(n)}_{nN-k_2}  S^{(n)}_{nN-k_3}S'^{(m')}_{m'N-k_4}  \ket{\Psi_{2N}}\notag\\
=& 2 V^k_R\ket{\Psi_{2N+2}} ,
\end{align}\end{widetext}
where we have used \Eq{rere} in the first step, $a_{i_2}b_{i_1}a_{r}^\dagger b_{k}^\dagger=-\delta_{r,i_2}\delta_{k,i_1}+\delta_{k,i_1} a_{i_2} a_{r}^\dagger + \delta_{r,i_2}b_{i_1} b_{k}^\dagger +a_{r}^\dagger b_{k}^\dagger  a_{i_2} b_{i_1}$ in the second step,  Eqs. \eqref{cb}, \eqref{arpsi}, \eqref{brpsi} and  $V^k_R S^{(m)}_{mN-k_1} S'^{(n)}_{nN-i_1+n-l+k_1}  S^{(n)}_{nN-k_3}S'^{(m')}_{m'N-k_4}  \ket{\Psi_{2N}}=0$ in the last step. 

We thus obtain $(N+1)V^k_R\ket{\Psi_{2N+2}}=2V^k_R\ket{\Psi_{2N+2}}$.
Therefore, if $\ket{\Psi_{2N}}$ is a zero mode of all  $V^k_R$ with  $2N  \geqslant  4 $, so will be $\ket{\Psi_{2N+2}}$. By mathematical induction, the second-quantized Halperin state, as recursively defined in \Eq{rere}, also satisfies the last zero mode condition \Eq{zo3}.\hfill $\blacksquare$

\subsection{Root state and filling factor of fermionic (bosonic) Halperin state}\label{RS}
We first introduce the notion of root state. In the expansion of a general FQH state $ \ket{\psi}$ in terms of occupation number basis states, there is a root state  $\ket{\psi}_{\text{root}}$ that is non-expandable in the sense that it cannot be obtained from any other basis state $\ket{\{n_i\}}$ via two-body inward-squeezing\cite{BH1, BH2, BH2.5, BH3, regnault, ortiz, Chen17,PartonFQH}. Explicitly,
\be
    \ket{\psi} = C_{\text{root}} \ket{\psi}_{\text{root}} +\sum_{\ket{\{n_i\}}\neq \ket{\psi}_{\text{root}}}
    C_{\{n_i\}} \ket{\{n_i\}}
\ee
with $C$ the respective expansion coefficients. Here, inward-squeezing \cite{BH2} is defined as any of the following operations $ a^{\dagger}_{j} a^{\dagger}_{i} a_{i-m} a_{j+m}$, $ b^{\dagger}_{j} b^{\dagger}_{i} b_{i-m} b_{j+m}$, $ a^{\dagger}_{j} b^{\dagger}_{i} b_{i-m} a_{j+m}$, and $ b^{\dagger}_{j} a^{\dagger}_{i} a_{i-m} b_{j+m}$ with $i \leqslant j$ and  $m > 0$.

The above definition of root state along with the conservation of orbitals (the angular momentum on disk and sphere, momentum on cylinder) also indicates that any basis state other than the root state can be obtained from the latter by a series of inward squeezing. The filling factor of a FQH state can be defined through root state,  
as $\nu=\Delta N / \Delta r_{\text{max}}$, where $\Delta N$ is the change in the particle number and $\Delta r_{\text{max}}$ is the change in the maximum occupied orbital in the root state.

In this subsection,  we will prove that the recursively defined fermionic (bosonic) Halperin state $\ket{\Psi_{2N}}$ in \Eq{rere} has a nonzero root state 
\begin{align} &(a_{n}^\dagger  b_{0}^\dagger +(-1)^n a_{0}^\dagger  b_{n}^\dagger)\notag
\\\times & (a_{m+2n}^\dagger  b_{m'+n}^\dagger +(-1)^n a_{m+n}^\dagger  b_{m'+2n}^\dagger)\notag
\\\times &\cdots \notag
\\\times & (a_{m(N-1)+nN}^\dagger  b_{m'(N-1)+n(N-1)}^\dagger\notag
\\& +(-1)^n a_{m(N-1)+n(N-1)}^\dagger  b_{m'(N-1)+nN}^\dagger)\ket{0}.   \end{align}
Again, we prove this by mathematical induction. For $N=1$, the above statement is true as seen from \Eq{psi2}.
Now we assume 
\begin{align}
&\ket{\Psi_{2N}}_{\text{root}}\notag\\\propto & \prod\limits_{j=1}^N \sum\limits_{l_j=0,n}(-1)^{l_j} {n \choose l_j}a_{(m+n)(j-1)+n-l_j}^\dagger  b_{(m'+n)(j-1)+l_j}^\dagger \ket{0} \end{align}
 for $N\geqslant 1$ and its coefficient in the expansion of $ \ket{\Psi_{2N}}$  in terms of occupation number basis states is nonzero.

We input $\ket{\Psi_{2N}}_{\text{root}}$ into \Eq{rere} to obtain
\begin{widetext}
\begin{align}
 & \frac{1}{N+1}\sum\limits_{l=0}^{n}(-1)^l {n \choose l} \sum_{k_1,k_2,k_3,k_4\geqslant 0}\, a_{n-l+k_1+k_2}^\dagger b_{l+k_3+k_4}^\dagger  S^{(m)}_{mN-k_1} S'^{(n)}_{nN-k_2}  S^{(n)}_{nN-k_3}S'^{(m')}_{m'N-k_4} \ket{\Psi_{2N}}_{\text{root}}\notag\\
= & \frac{1}{N+1}\sum\limits_{l=0}^{n}(-1)^l {n \choose l} \sum_{k_1,k_2,k_3,k_4\geqslant 0}\, a_{n-l+k_1+k_2}^\dagger b_{l+k_3+k_4}^\dagger  S^{(m)}_{mN-k_1} S'^{(n)}_{nN-k_2}  S^{(n)}_{nN-k_3}S'^{(m')}_{m'N-k_4} \notag\\
 &\times \prod\limits_{j=1}^N \sum\limits_{l_j=0,n}(-1)^{l_j} {n \choose l_j}a_{(m+n)(j-1)+n-l_j}^\dagger  b_{(m'+n)(j-1)+l_j}^\dagger\ket{0} \notag
\\=  & \frac{1}{N+1}\sum\limits_{l=0}^{n}(-1)^l {n \choose l} \sum_{k_1,k_2,k_3,k_4\geqslant 0}\sum\limits_{p_1,\dots p_N=0}^{m}\sum\limits_{q_1,\dots q_N,r_1,\dots r_N=0}^{n}\sum\limits_{s_1,\dots s_N=0}^{m'}(-1)^{\sum_{i=1}^N (p_i+q_i+r_i+s_i)} \prod_{i=1}^N{m \choose p_i}{n \choose q_i}{n \choose r_i}{m' \choose s_i}\notag\\&\times   \, a_{n-l+k_1+k_2}^\dagger b_{l+k_3+k_4}^\dagger   \prod\limits_{j=1}^N \sum\limits_{l_j=0,n}(-1)^{l_j} {n \choose l_j}a_{(m+n)(j-1)+n-l_j+p_j+q_j}^\dagger  b_{(m'+n)(j-1)+l_j+r_j+s_j}^\dagger \notag\\
&\times  S^{(m)}_{mN-k_1-\sum_{i=1}^N p_i} S'^{(n)}_{nN-k_2-\sum_{i=1}^N r_i}  S^{(n)}_{nN-k_3-\sum_{i=1}^N q_i}S'^{(m')}_{m'N-k_4-\sum_{i=1}^N s_i}\ket{0} \notag\\
=  & \frac{1}{N+1}\sum\limits_{l_{N+1}=0}^{n}(-1)^{l_{N+1}} {n \choose l_{N+1}} \sum\limits_{p_1,\dots p_N=0}^{m}\sum\limits_{q_1,\dots q_N,r_1,\dots r_N=0}^{n}\sum\limits_{s_1,\dots s_N=0}^{m'}(-1)^{\sum_{i=1}^N (p_i+q_i+r_i+s_i)} \prod_{i=1}^N{m \choose p_i}{n \choose q_i}{n \choose r_i}{m' \choose s_i}\notag\\&\times   \, a_{n-l_{N+1}+(m+n)N-\sum_{i=1}^N (p_i+r_i)}^\dagger b_{l_{N+1}+(m'+n)N-\sum_{i=1}^N (q_i+s_i)}^\dagger  \notag\\
  & \times \prod\limits_{j=1}^N \sum\limits_{l_j=0,n}(-1)^{l_j} {n \choose l_j}a_{(m+n)(j-1)+n-l_j+p_j+q_j}^\dagger  b_{(m'+n)(j-1)+l_j+r_j+s_j}^\dagger  \ket{0},\notag
\end{align}\end{widetext}
where we have used Eqs. \eqref{Sad} and \eqref{Sbd} to move $S$ and $S'$ to the right of $a^\dagger$ and $b^\dagger$. In the above derivation, we have also used the fact that $S^{(m)}_{k}\ket{0}$ and  $S'^{(m')}_{k}\ket{0}$ vanish unless $k=0$, as in the derivation of \Eq{psi2}.

To generate $\ket{\Psi_{2N+2}}_{\text{root}}$, $a_{n-l_{N+1}+(m+n)N-\sum_{i=1}^N (p_i+r_i)}^\dagger b_{l_{N+1}+(m'+n)N-\sum_{i=1}^N (q_i+s_i)}^\dagger $ in the above equation should be equal to $a_{(m+n)(k-1)+n-l_k}^\dagger  b_{(m'+n)(k-1)+l_k}^\dagger$ with $1\leqslant k\leqslant N+1$, while \begin{align}&\prod\limits_{j=1}^N \sum\limits_{l_j=0,n}(-1)^{l_j} {n \choose l_j}\notag\\\times & a_{(m+n)(j-1)+n-l_j+p_j+q_j}^\dagger  b_{(m'+n)(j-1)+l_j+r_j+s_j}^\dagger \end{align} are equal to other terms in $\ket{\Psi_{2N+2}}_{\text{root}}$.

We find the solution for the given $k$ is  $ p_1=q_1=r_1=s_1=\dots= p_{k-1}= q_{k-1}= r_{k-1}= s_{k-1}=0$, $ p_{k}=p_{k+1}=\dots= p_{N}=m$, $ q_{k}= r_{k}=q_{k+1}=r_{k+1}=\dots= q_{N}= r_{N}=n$,  $ s_{k}=s_{k+1}=\dots=s_{N}=m'$, and $l_i=0,n$ for $1 \leqslant i \leqslant N+1$.

As a consequence, $\ket{\Psi_{2N+2}}_{\text{root}}$ can be generated from $\ket{\Psi_{2N}}_{\text{root}}$ via \Eq{rere}.
On the other hand,  $\ket{\{n_i\}}$ other than $\ket{\Psi_{2N}}_{\text{root}}$ in  the expansion of  $\ket{\Psi_{2N}}$ cannot generate $\ket{\Psi_{2N+2}}_{\text{root}}$. 

In conclusion, fermionic (bosonic) Halperin state $\ket{\Psi_{2N}}$, as recursively defined in \Eq{rere},  has a non-vanishing root state 
\begin{align}
 & \prod\limits_{j=1}^N \sum\limits_{l_j=0,n}(-1)^{l_j} {n \choose l_j}a_{(m+n)(j-1)+n-l_j}^\dagger  b_{(m'+n)(j-1)+l_j}^\dagger \ket{0}. \end{align}
By definition, the filling factor for the two layers are $1/(m+n)$ and $1/(m'+n)$, respectively. If $m'=m$, the total filling factor would be $2/(m+n)$. All these agrees with the filling factors obtained from the first-quantized wave function.

Furthermore, If we regard the layer degree of freedom as (pseudo) spin degree of freedom, hence rewrite $a_{k}^\dagger$ and  $b_{k}^\dagger$ as $c_{k,\uparrow}^\dagger$ and  $c_{k,\downarrow}^\dagger$, respectively, the root state will take the form \begin{align} &(c_{n,\uparrow}^\dagger  c_{0,\downarrow}^\dagger +(-1)^n c_{0,\uparrow}^\dagger  c_{n,\downarrow}^\dagger)\notag
\\\times & (c_{m+2n,\uparrow}^\dagger  c_{m'+n,\downarrow}^\dagger +(-1)^n c_{m+n,\uparrow}^\dagger  c_{m'+2n,\downarrow}^\dagger)\notag
\\\times &\cdots \notag
\\\times & (c_{m(N-1)+nN,\uparrow}^\dagger  c_{m'(N-1)+n(N-1),\downarrow}^\dagger\notag
\\& +(-1)^n c_{m(N-1)+n(N-1),\uparrow}^\dagger  c_{m'(N-1)+nN,\downarrow}^\dagger)\ket{0}.   \end{align}
In the case $m'=m$, it is obvious that the root state has the structure of product of spin-singlets if $n$ is odd and spin-triplets if $n$ is even. This reflects the SU(2) symmetry of Halperin $(m,m,n)$ state.

\section{Conclusion and Discussion}\label{condisc}
Based on our previous work in the second-quantized approach to Laughlin state, Jain composite fermion state and Pfaffian state, in this paper we generalize this approach to multi-component FQH state, specifically the Halperin $(m,m',n)$ state.
In a self-consistent manner, we define the second-quantized Halperin state recursively in particle number via a recursion relation \Eq{rere}.
We rigourously prove that the Halperin state, as recursively defined in \eqref{rere},  is indeed a zero mode of the corresponding second-quantized parent Hamiltonian \eqref{Ham} and that it has the correct filling factor. We also demonstrate that in the case $m'=m$, the root state of the second-quantized Halperin state manifests the structure of SU(2) symmetry. Our second-quantized approach thus complements the first-quantized approach taken by other researchers in the field.


This approach can be generalized to other multi-component FQH states, e.g. the Read-Rezayi state\cite{RR}. 
The generalization to cluster FQH state,  for example the Gaffnian state\cite{gaffnian},  is also viable, but needs special care. This state also has a two-component structure as the Halperin state, but its wave function has a (anti-) symmetrization operator in front of the two-component part (depending on whether the constituting particles are fermions or bosons), rendering it single component. In the construction of recursion relation for the Gaffnian state, we can first construct recursion relation for  the Gaffnian state \textit{without} the (anti-) symmetrization, whose form will resemble \Eq{rere}. The obtained second-quantized Gaffnian state of particle number $2N$ \textit{without} the (anti-) symmetrization will have an equal number of $a^\dag$ and $b^\dag$. In the next step, the (anti-) symmetrization operation is enforced by letting $b^\dag=a^\dag$, and then using (anti-) commutation relations for (fermionic) bosonic $a^\dag$. This will be presented in a  work in the future\cite{Gaffrecur}.

\begin{acknowledgments}
Li Chen acknowledges support  by NSFC Grant No. 12004105.\end{acknowledgments}
\appendix
\begin{widetext}

\section{The derivation of \Eq{rere} from the first-quantized wave function}\label{1stTo2nd}
We can write down the wave function of $2N$-particle Halperin $(m,m',n)$ state as $\Psi_{2N}$. When the particle number increases by two, we have the relation:
\begin{align}
\Psi_{2N+2} &=(z_{N+1}-w_{N+1})^n \prod\limits_{i=1}^{N}[(z_{N+1}-z_{i})^m (z_{N+1}-w_{i})^n (w_{N+1}-z_{i})^n (w_{N+1}-w_{i})^{m'}]\Psi_{2N}.\end{align}

In the first place, we need to expand $\prod\limits_{i=1}^{N}(z_{N+1}-z_{i})^m$. For this purpose, we  expand
\begin{align}\prod\limits_{i=1}^{N}(z_{N+1}-z_i)=&\sum\limits_{k=0}^N  z_{N+1}^k (-1)^{N-k}\sum\limits_{i_1<i_2\cdots<i_{N-k}} z_{i_1}z_{i_2}\cdots z_{i_{N-k}}\notag\\=&\sum\limits_{k=0}^N  z_{N+1}^k (-1)^{N-k}e_{N-k},\end{align} where we have defined  elementary-symmetric-polynomial operator $e_{N-k}=\sum\limits_{ i_1<i_2\cdots<i_{N-k}} z_{i_1}z_{i_2}\cdots z_{i_{N-k}}$. Its second-quantized form is defined in \Eq{e}.

Then we have \be\prod\limits_{1\leqslant i\leqslant N}(z_{N+1}-z_i)^m=\sum\limits_{k=0}^{mN}  z_{N+1}^k S^{(m)}_{mN-k}, \ee where $S$ is related to $e$ by \Eq{Sm}. Likewise, we obtain  \be\prod\limits_{1\leqslant i\leqslant N}(w_{N+1}-z_i)^n=\sum\limits_{k=0}^{nN}  w_{N+1}^k S^{(n)}_{nN-k}, \ee
\be\prod\limits_{1\leqslant i\leqslant N}(z_{N+1}-w_i)^n=\sum\limits_{k=0}^{nN}  z_{N+1}^k S'^{(n)}_{nN-k}, \ee
\be\prod\limits_{1\leqslant i\leqslant N}(w_{N+1}-w_i)^{m'}=\sum\limits_{k=0}^{m'N}  w_{N+1}^k S'^{(m')}_{m'N-k}, \ee where $S'$ is defined for the second layer, in the same way as $S$ is defined for the first layer. 
Finally, $(z_{N+1}-w_{N+1})^{n}$ can be expanded via binomial expansion. We hence obtain
\be\label{nplus2}\Psi_{2N+2}=\sum_{k_1,k_2,k_3,k_4}\sum\limits_{l=0}^{n}(-1)^l {n \choose l}z_{N+1}^{n-l+k_1+k_2}w_{N+1}^{l+k_3+k_4}S^{(m)}_{mN-k_1} S'^{(n)}_{nN-k_2}  S^{(n)}_{nN-k_3}S'^{(m')}_{m'N-k_4} \Psi_{2N}.\ee

Upon second quantization, the above formula leads to
\be\label{reforAppend}
\ket{\Psi_{2N+2}}= \frac{1}{N+1}\sum\limits_{l=0}^{n}(-1)^l {n \choose l} \sum_{k_1,k_2,k_3,k_4}\, a_{n-l+k_1+k_2}^\dagger b_{l+k_3+k_4}^\dagger  S^{(m)}_{mN-k_1} S'^{(n)}_{nN-k_2}  S^{(n)}_{nN-k_3}S'^{(m')}_{m'N-k_4}  \ket{\Psi_{2N}}, \ee  the prefactor comes from  Eq. (1.13) of Ref. \onlinecite{cr2nd} with each layer having $N+1$ particles.

Employing both Eq.~(1.12) and Eq.~(1.13) of Ref. \onlinecite{cr2nd},
we can obtain the expression for  Halperin state with one particle removed from the first or second layer
\begin{align}\label{arpsi}
a_r \ket{\Psi_{2N+2}}=& \sum_{k_1,k_2,k_3,k_4}  (-1)^{n-r+k_1+k_2} {n \choose r-k_1-k_2} \,  b_{n-r+k_1+k_2+k_3+k_4}^\dagger  \notag\\&\times S^{(m)}_{mN-k_1} S'^{(n)}_{nN-k_2}  S^{(n)}_{nN-k_3}S'^{(m')}_{m'N-k_4}  \ket{\Psi_{2N}},\end{align}
and \begin{align}\label{brpsi}
b_r \ket{\Psi_{2N+2}}=& \sum_{k_1,k_2,k_3,k_4}(-1)^{r-k_3-k_4} {n \choose r-k_3-k_4}\, a_{n-r+k_1+k_2+k_3+k_4}^\dagger  \notag\\&\times S^{(m)}_{mN-k_1} S'^{(n)}_{nN-k_2}  S^{(n)}_{nN-k_3}S'^{(m')}_{m'N-k_4}  \ket{\Psi_{2N}}.\end{align}
Note than the above two identities can also lead back to \Eq{reforAppend} by using the fact that $\ket{\Psi_{2N+2}}$ is the eigenstate of the particle number operator in either layer with the eigenvalue $N+1$:
\be \ket{\Psi_{2N+2}}= \frac{1}{N+1}\sum_{r}  a_r^\dagger a_r \ket{\Psi_{2N+2}}=\frac{1}{N+1}\sum_{r}  b_r^\dagger b_r \ket{\Psi_{2N+2}}.\ee

\end{widetext}

\bibliography{Halp}
\end{document}